\documentclass[iop,preprint2]{emulateapj1}
\usepackage{psfig}
\usepackage{apjfonts}
\usepackage{ulem}
\usepackage[dvips]{color}
\usepackage{graphicx}
\usepackage{rotating}
\usepackage[USenglish]{babel}
\usepackage{mathrsfs}
\usepackage{longtable}

\newcommand{\wcen}{$\omega$Cen}
\newcommand{\ebv}{E($B-V$)}

\begin{document}

\title{THE STATE-OF-THE-ART \textit{HST} ASTRO-PHOTOMETRIC ANALYSIS OF
  THE CORE OF $\omega$~CENTAURI. II. DIFFERENTIAL-REDDENING
  MAP$^{\ast}$}

\author{A.\ Bellini\altaffilmark{1},
J.\ Anderson\altaffilmark{1},
R.\ P.\ van der Marel\altaffilmark{1},
I.\ R.\ King\altaffilmark{2},
G.\ Piotto\altaffilmark{3,4},
and L.\ R.\ Bedin\altaffilmark{4}}

\altaffiltext{1}{Space Telescope Science Institute, 3700 San Martin
  Dr., Baltimore, MD 21218, USA}
\altaffiltext{2}{Department of Astronomy, University of Washington,
  Box 351580, Seattle, 98195, WA, USA}
\altaffiltext{3}{Dipartimento di Fisica e Astronomia ``Galileo
  Galilei'', Universit\`a di Padova, Vicolo dell'Osservatorio 3,
  Padova I-35122, Italy}
\altaffiltext{4}{Istituto Nazionale di Astrofisica, Osservatorio
  Astronomico di Padova, v.co dell'Osservatorio 5, I-35122, Padova,
  Italy}
\altaffiltext{$^\ast$}{Based on archival observations with the
  NASA/ESA \textit{Hubble Space Telescope}, obtained at the Space
  Telescope Science Institute, which is operated by AURA, Inc., under
  NASA contract NAS 5-26555.}
\received{February 14, 2014}
\revised{April 14, 2017}
\accepted{April 18, 2017}

\email{bellini@stsci.edu}

\begin{abstract} {
We take advantage of the exquisite quality of the \textit{Hubble Space
  Telescope} astro-photometric catalog of the core of \wcen\ presented
in the first paper of this series to derive a high-resolution,
high-precision, high-accuracy differential-reddening map of the
field. The map has a spatial resolution of 2$\times$2 arcsec$^2$ over
a total field of view of about $4\farcm3$$\times$$4\farcm3$.  The
differential reddening itself is estimated via an iterative procedure
using five distinct color-magnitude diagrams, which provided
consistent results to within the 0.1\% level.  Assuming an average
reddening value E($B-V$)=0.12, the differential-reddening within the
cluster's core can vary by up to $\pm$10\%, with a typical a standard
deviation of about 4\%.  Our differential-reddening map is made
available to the astronomical community in the form of a
multi-extension \texttt{FITS} file.  This differential-reddening map
is essential for a detailed understanding of the multiple stellar
populations of \wcen, as presented in the next paper in this
series. Moreover, it provides unique insight into the level of small
spatial-scale extinction variations in the Galactic foreground.}
\end{abstract}

\keywords{
globular clusters: individual (NGC 5139) ---
Hertzsprung-Russell and C-M diagrams --- 
stars: Population II --- 
techniques: photometric}

\maketitle

\section{Introduction}
\label{s:intro}

The globular cluster (GC) NGC~5139 (\wcen) is one of the best-studied
objects in the sky. It is the most massive GC of the Milky Way
(4.55$\times$10$^6$ $M_{\odot}$, \citealt{2013MNRAS.429.1887D}), and
its relative proximity to the Sun ($\sim$5.2 kpc,
\citealt{1996AJ....112.1487H}) has made it an excellent target for a
large variety of photometric and spectroscopic investigations over the
past 100 years.

\wcen\ is affected by only a mild reddening (\ebv=0.12,
\citealt{1996AJ....112.1487H}). \cite{2005ApJ...634L..69C}, using a
technique based on the color excess of hot horizontal-branch stars,
found sizable clumpy extinction variations of up to a factor of two in
the cluster's central regions.  Large differential-reddening (DR)
variations in the core of \wcen\ has been disputed in the
literature. For instance, \cite{2007ApJ...663..296V}, using
\textit{Hubble Space Telescope} (\textit{HST}) observations of the
core of the cluster taken with the Wide-Field Channel (WFC) of the
Advanced Camera for Surveys (ACS), pointed out that, given the
sharpness of the cluster's sequences on a color-magnitude diagram
(CMD, see, e.g., their Figures 1--6), the existence of any serious
differential reddening is very unlikely. The authors did not
correct their photometry for DR, nor did they attempt to derive a DR
map of their field, and therefore their statement is only qualitative
and not quantitative.

In \cite{2010AJ....140..631B}, we analyzed the core of \wcen\ using
the very first (few) \textit{HST} UV photometric observations taken
with the Ultraviolet-VISible channel (UVIS) of the newly installed
Wide-Field Camera 3 (WFC3) during science verification after the
camera was installed in 2009. We applied a small correction to our
photometry and were able to minimize the effects of both reddening
variations and spatial-dependent photometric errors introduced by
small variations of the PSF shape (and unaccounted for by the PSF
models). The lack of an adequate number of exposures taken at
different roll angles and a non optimal dither pattern prevented us
from empirically separating the contribution of PSF-induced variations
from the DR itself. Nevertheless, if the PSF-induced variations had
been negligible, then the applied correction would have implied a DR
variation in the core of the cluster of up to $\pm 12$\% with respect
to the average reddening.

The core of \wcen\ has been chosen as a calibration field for the
WFC3, and the \textit{HST} archive now contains several hundreds of
exposures taken through many different filters, roll angles, and
epochs. To exploit this, we submitted an archival proposal (AR-12656,
PI:\ J.\ Anderson) to produce the most-comprehensive catalog of
photometry and proper-motions ever assembled for the cluster's
core. In \citet[hereafter Paper~I]{paper1}, we describe in
detail the data-reduction procedures we applied to obtain
high-precision photometry and astrometry for over 470\,000 stars
within the central $\sim$$2\farcm7$ (the cluster's core radius being
$2\farcm37$, \citealt{1996AJ....112.1487H}). Photometry is measured
through 26 filters, with a continuous wavelength coverage from the UV
(F225W) to the IR (F160W). Well-measured stars in the catalog have
typical proper-motion errors of about 25 $\mu$as yr$^{-1}$, or 0.6 km
s$^{-1}$ at a distance of 5.2 kpc.

In this second paper of the series, we focus our attention on deriving
a high-precision DR map of the field. The goal is twofold:\ on one
hand, DR-corrected photometry will greatly improve our ability to
select and analyze the multiple stellar populations hosted by the
cluster, which will be the topic of future papers in this series. On
the other hand, we want to determine once and for all the true amount
of DR in the core of the cluster.

This paper is organized as follows. Section~\ref{s:meth} describes the
methodology we adopted to correct for DR in minute detail. In
Section~\ref{s:maps} we critically analyze the DR map we obtain and
estimate the possible impact of systematic effects.  Discussion and
conclusions are presented in Section~\ref{s:disc}.

\section{Methodology}
\label{s:meth}

We closely followed the DR-correction techniques described in detail
in Section~3.1 of \cite{2012A&A...540A..16M} and in Section~2.5 of
\cite{2013ApJ...765...32B}.  In a nutshell, we start from the
reasonable assumption that cluster stars are all at the same distance
from us, and that the amount of intra-cluster DR is negligible. For a
given target star in the field of view (FoV), we measure a local
differential reddening as the median shift along the reddening vector
in a CMD of a set of reference cluster stars with respect to their
average fiducial line. Such deviation represents the amount of local
DR correction affecting the target star. This technique is designed to
correct only for the foreground contribution to DR.  The DR correction
itself is based on a set of well-measured, bright, unsaturated
reference stars.

The best candidate reference stars in a CMD lie on a sequence that is
as perpendicular as possible to the direction of the reddening vector
(typically along the sub-giant branch, SGB), i.e, where the broadening
of the sequence itself is most sensitive to DR effects.  If the chosen
sequence is made up of stars of similar chemical composition and age,
then in absence of DR the broadening of the sequence should only be a
direct consequence of photometric errors. Thus, the broader the
sequence with respect to what photometric errors predict, the larger
the amount of DR in the field is to be expected.

It is now well known that formally all GCs host multiple populations
of stars (e.g, \citealt{2015AJ....149...91P}), and \wcen\ is probably
the most famous and extreme example (e.g.,
\citealt{2004ApJ...605L.125B, 2010AJ....140..631B}). The SGB of
\wcen\ is a tangled twining of sequences made of stars of different
iron, light-element, and helium abundances (e.g.,
\citealt{2011ApJ...731...64M, 2005ApJ...621..777P,
  2012AJ....144....5K}), and possibly of different age
(\citealt{2007ApJ...663..296V, 2014ApJ...791..107V}).

In principle, it is always possible to choose the SGB in its entirety
as a reference to correct for DR effects.  This choice, anyway, leads
to somewhat large uncertainties in the achievable DR correction, since
the correction accuracy is proportional to the intrinsic dispersion of
reference stars along the reddening vector. Therefore, the narrower
the intrinsic sequence of reference stars, the more accurate the DR
correction. Of course, in the case of \wcen, we need to be extra
careful in selecting reference stars belonging to the same sequence,
due to the presence of numerous stellar populations, and this is
especially true when the populations overlap each other, as it is the
case for the SGB of \wcen.

The FoV-averaged fiducial line defined by the reference stars
constitutes the baseline with respect to which we compute the DR
correction. This choice guarantees us to have a net DR correction of
zero over the entire FoV. Then, for each star in the catalog, the
associated DR correction is simply obtained as the median displacement
of the nearest $n_\ast$ reference stars along the direction of
reddening vector with respect to the fiducial line. The number of
reference stars is chosen by trial and error as a compromise between
a small $n_\ast$ for spatial resolution, and a large $n_\ast$ for
robustness. We found that $n_\ast$=75 works well for the FoV and data
set at our disposal.

In general, this DR-correction procedure also minimizes small
PSF-related zero-point variations across the FoV, as was the case
for the \cite{2010AJ....140..631B} work.  There can be various reasons
why a PSF model might not be ``perfect''. To start, it is well known that
the shape of the \textit{HST}'s PSF varies spatially by about 10\%
across the detector.

Moreover, due to telescope breathing, temporal variations of similar
amplitude are present from one exposure to the next. The PSF models we
employed in paper~I accurately reproduce both spatial and temporal
variations (see, e.g., their Section~3.1). However, small systematic
residuals might still be present in our high-precision photometry. One
simple way to quantify the PSF-related contribution to the DR
correction is to compare DR maps obtained by using CMDs based on
different filters.  To the first order, the average of these maps is
expected to be free of PSF-related systematic effects, while the
difference between the average DR map and the individual,
filter-dependent maps provides a direct estimate of PSF-related
systematic effects (more in Sect.~\ref{s:maps}).

To this aim, we computed five DR maps based on CMDs employing five
representative WFC3/UVIS filters: F275W, F336W, F438W, F606W and
F814W. These five filters offer the largest FoVs in our data set, and
their numerous exposures were taken at different roll angles and
dithers, thus minimizing --to the extent possible-- PSF-related
systematic errors. We constructed five CMDs with fixed F275W$-$F814W
color, i.e., the widest, and we let the magnitude vary each time, from
F275W to F814W. The wide F275W$-$F814W CMD color helps in keeping the
reddening vector's angle of incidence over the SGB sequences as
orthogonal as possible, thus maximizing the DR contribution to the
broadening of the SGB sequences.

In the following, we will describe in detail each step of the
DR-correction process.

\begin{figure*}[!t]
\centering
\includegraphics[width=\textwidth]{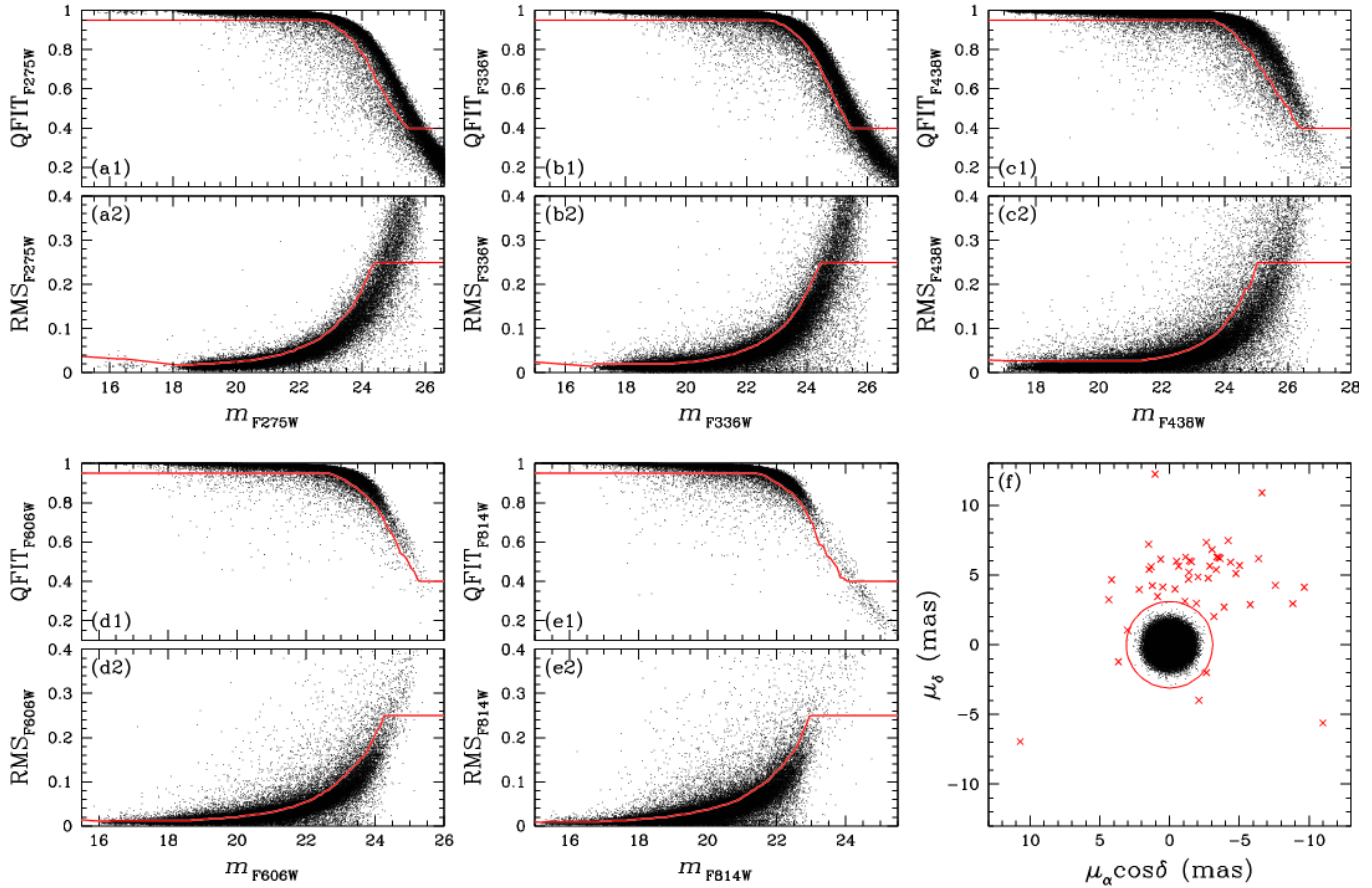}
\caption{\small{Selection criteria applied to the \texttt{QFIT} and
    photometric RMS parameters of the five adopted filters (panels
    a1-2, b1-2, c1-2, d1-2, and e1-2). In each of these 10 panels, we
    plotted only 30\% randomly-selected stars, for clarity. The red
    curve in each panel separated the bulk of well-measured stars from
    the few outliers.  Stars that appear to be good according to all
    selection criteria are plotted in the proper-motion diagram of
    panel (f), where we can further refine our selections by rejecting
    the few field stars (red crosses outside the circle) present in
    our FoV. (See the text for details.)}}
\label{f:sel}
\end{figure*}

\subsection{Extinction Coefficients}
\label{ss:extc}

The extinction coefficients for the five representative WFC3/UVIS
filters were obtained using the York Extinction Solver
(YES)\footnote[1]{http://www.cadc-ccda.hia-iha.nrc-cnrc.gc.ca/community/\\ YorkExtinctionSolver/}
website. We adopted the \cite{1999PASP..111...63F} extinction law,
scaled to R$_V$=3.1, and calculated the extinction coefficients at the
reference wavelength of each filter.  These values are reported in
column 2 of Table~1.  We assumed an average reddening over our FoV to
be E($B$$-$$V$)=0.12
(\citealt{1996AJ....112.1487H}).\footnote[2]{Please note that we will
  measure the amount of DR with respect to the reference-star fiducial
  line defined over the entire FoV. As a consequence, the DR maps we
  will obtain are intended as relative DR maps, and are independent
  from the true value of E($B$$-$$V$).}

\subsection{Sample Selection}
\label{ss:sel}

\begin{figure*}[!t]
\centering
\includegraphics[width=\textwidth]{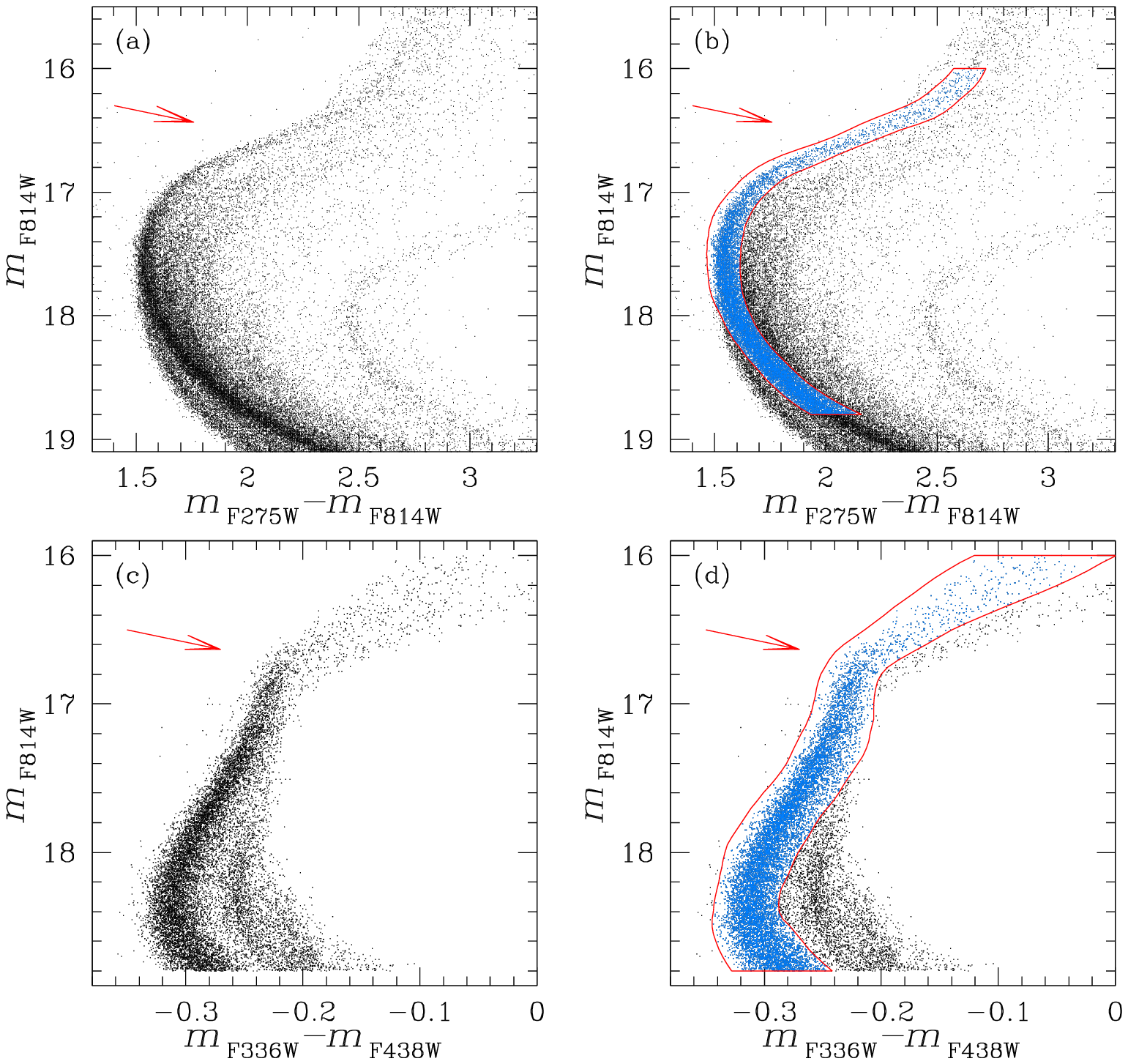}
\caption{\small{(a) The $m_{\rm F814W}$ vs.$m_{\rm F275W}$-$m_{\rm
      F814W}$ CMD of \wcen\ around the SGB region.  Panel (b) is a
    replica of panel (a), in which we initially selected a sample of
    reference stars that lie on the most populated sequence of the
    cluster (i.e., the rMS and its SGB progeny). Selected stars
    are in azure within the red box.  Panel (c) shows the $m_{\rm
      F814W}$ vs.\ $m_{\rm F336W}$-$m_{\rm F438W}$ CMD of stars as
    selected in panel (b).  It is known that the blue and the red MSs
    of \wcen\ flip with respect to each other in a CMD based on this
    particular color (see, e.g., \citealt{2010AJ....140..631B}). We
    exploited this fact to further refine our reference-star
    selection. Panel (d) is a replica of panel (c) in which we removed
    likely blue-MS contaminants from our reference-star sample. The
    final sample of reference stars is shown in azure within the red
    selection box.  The red arrow in all four panels highlights the
    direction of the reddening vector.}}
\label{f:ref}
\end{figure*}

\begin{figure*}[!t]
\centering
\includegraphics[width=\textwidth]{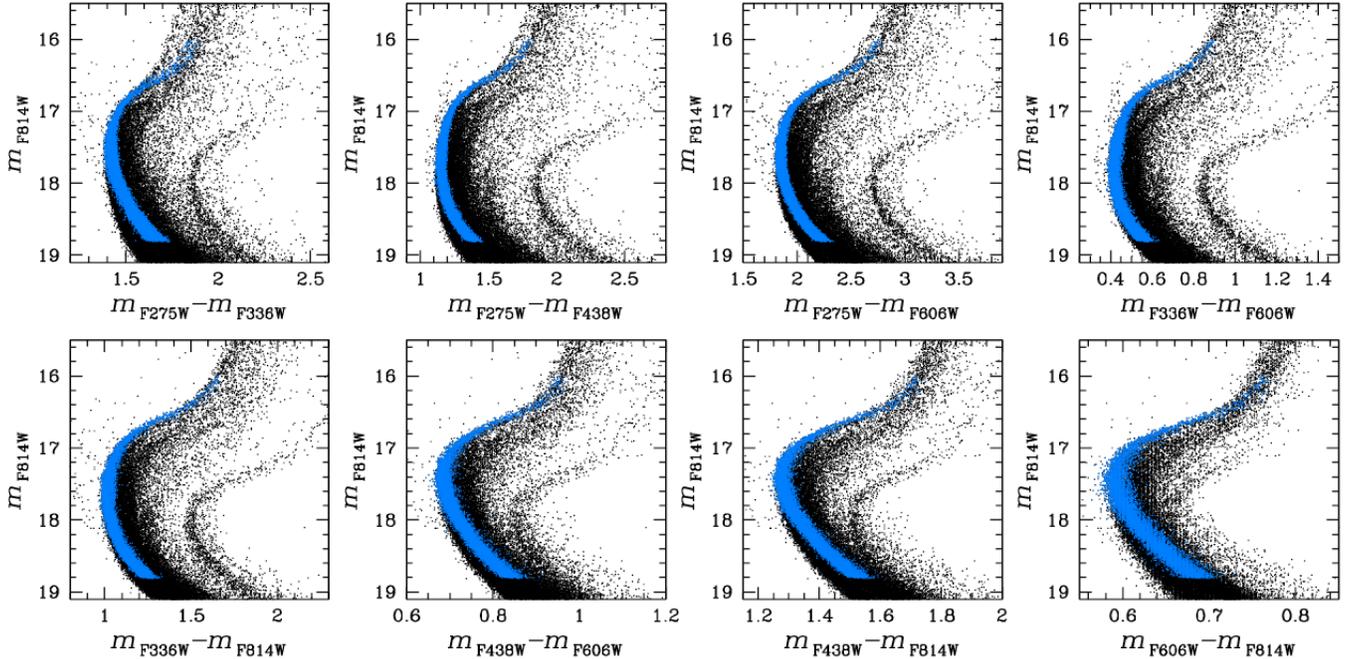}
\caption{\small{These CMDs, which color is made up by all
      possible filter combinations, show that reference stars define a
      single sequences in all cases. We are omitting here the CMDs
      based on $m_{\rm F275W}-m_{\rm F814W}$ and $m_{\rm F336W}-m_{\rm
        F438W}$ colors, since they already present in
      Fig.~\ref{f:ref}.}}
\label{f:show}
\end{figure*}

\begin{figure*}[!t]
\centering
\includegraphics[width=\textwidth]{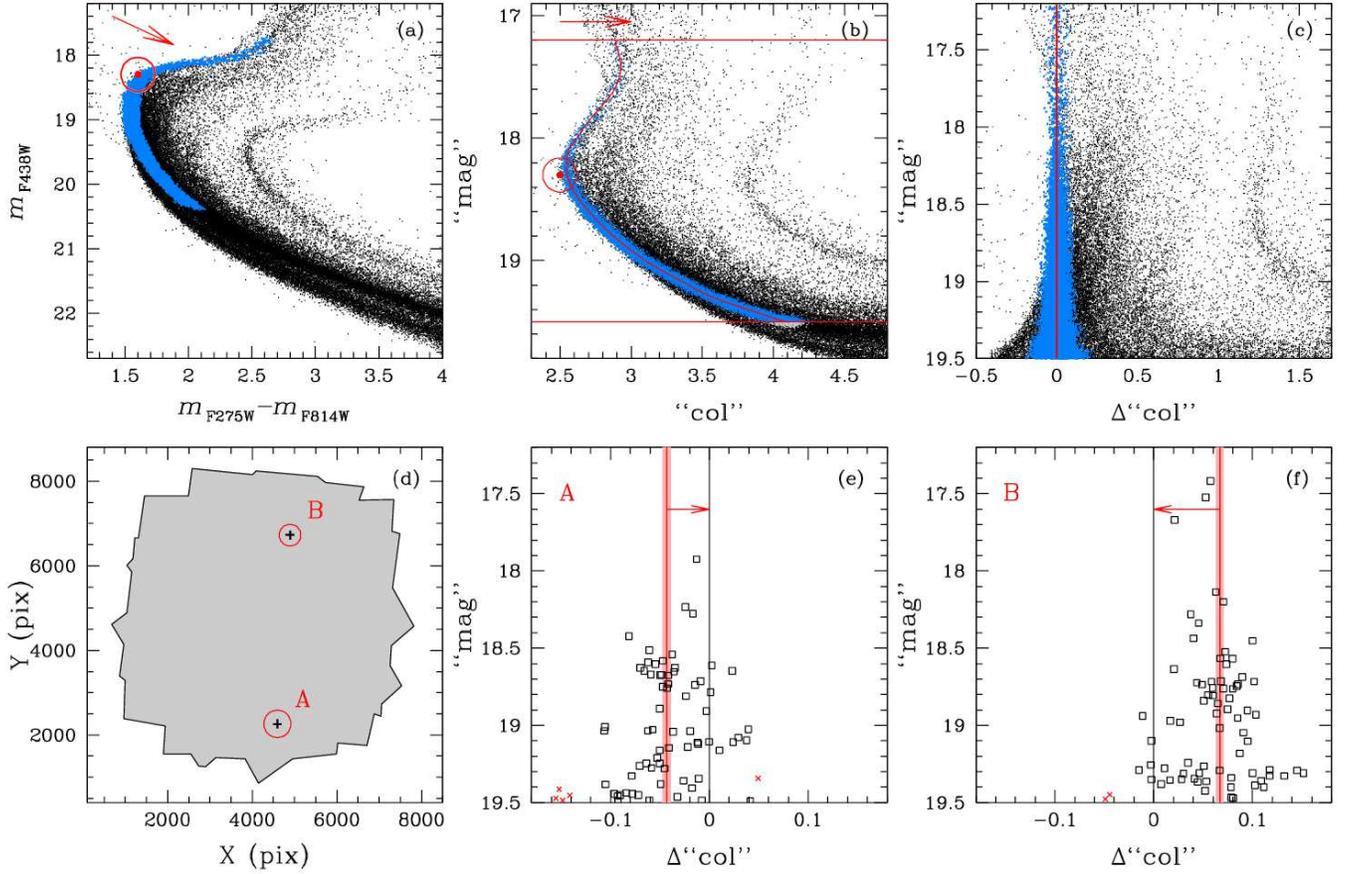}
\caption{\small{This figure illustrates how our DR-correction routine
    works. (a) the $m_{\rm F438W}$ vs.$m_{\rm F275W}$-$m_{\rm F814W}$
    CMD with the reference stars highlighted in azure and the
    reddening vector shown in red. We chose an arbitrary point (red
    encircled dot) around which we rotated the CMD in such a way that
    the reddening vector direction becomes parallel to the abscissa.
    The resulting, rotated pseudo-CMD is in panel (b). We call the
    abscissa and ordinate of this plane as ``col'' and ``mag'',
    respectively.  The reddening vector, now horizontal, is shown in
    red, for completeness, as well as the location of the rotation
    point.  On this plane, we derived a fiducial line (red curve)
    using the reference stars (in azure). The fiducial line is then
    used to rectify the reference-star sequence.  In some cases, at
    the bright and/or the faint end, reference stars do not equally
    populate both sides of the fiducial line. These stars (shown here
    in gray at the faint end) are excluded from the analysis to avoid
    DR-correction biases. The rectification is done by subtracting
    from the ``col'' value of each star that of the fiducial line at
    the same ``mag'' level. The rectified pseudo-CMD is shown in panel
    (c). We call the abscissa of the rectified pseudo-CMD as
    $\Delta$``col''.  Panel (d) shows the footprint of the FoV, in
    which we marked the location of two target stars, A and B, which
    are used as an example to illustrate how their DR correction is
    computed. We identified the 75 nearest reference stars to
    locations A and B on the field, which happened to be enclosed
    within the two red circles in panel (d). Panels (e) and (f) show
    the ``mag'' vs. $\Delta$``col'' pseudo-CMDs of the closest 75
    reference stars to A and B, respectively.  The 2.5$\sigma$-clipped
    median $\Delta$``col'' value of the $\Delta$``col'' of these
    reference stars (red vertical lines in panels e and f) is then
    used to correct the $\Delta$``col'' value of the target stars, and
    represent our best estimate of the local amount of DR.  The
    associated $\pm$ standard error of the mean is illustrated by a
    pink vertical band. In both cases, the $\Delta$ ``col'' standard
    error is about 0.0042, which translates into a magnitude and color
    error of about 0.003 magnitudes. Rejected reference stars are
    shown as red crosses in both panels (e) and (f). The applied
    correction in both cases is shown as a red horizontal arrow.}}
\label{f:how}
\end{figure*}

\begin{figure*}[!t]
\centering
\includegraphics[width=\textwidth]{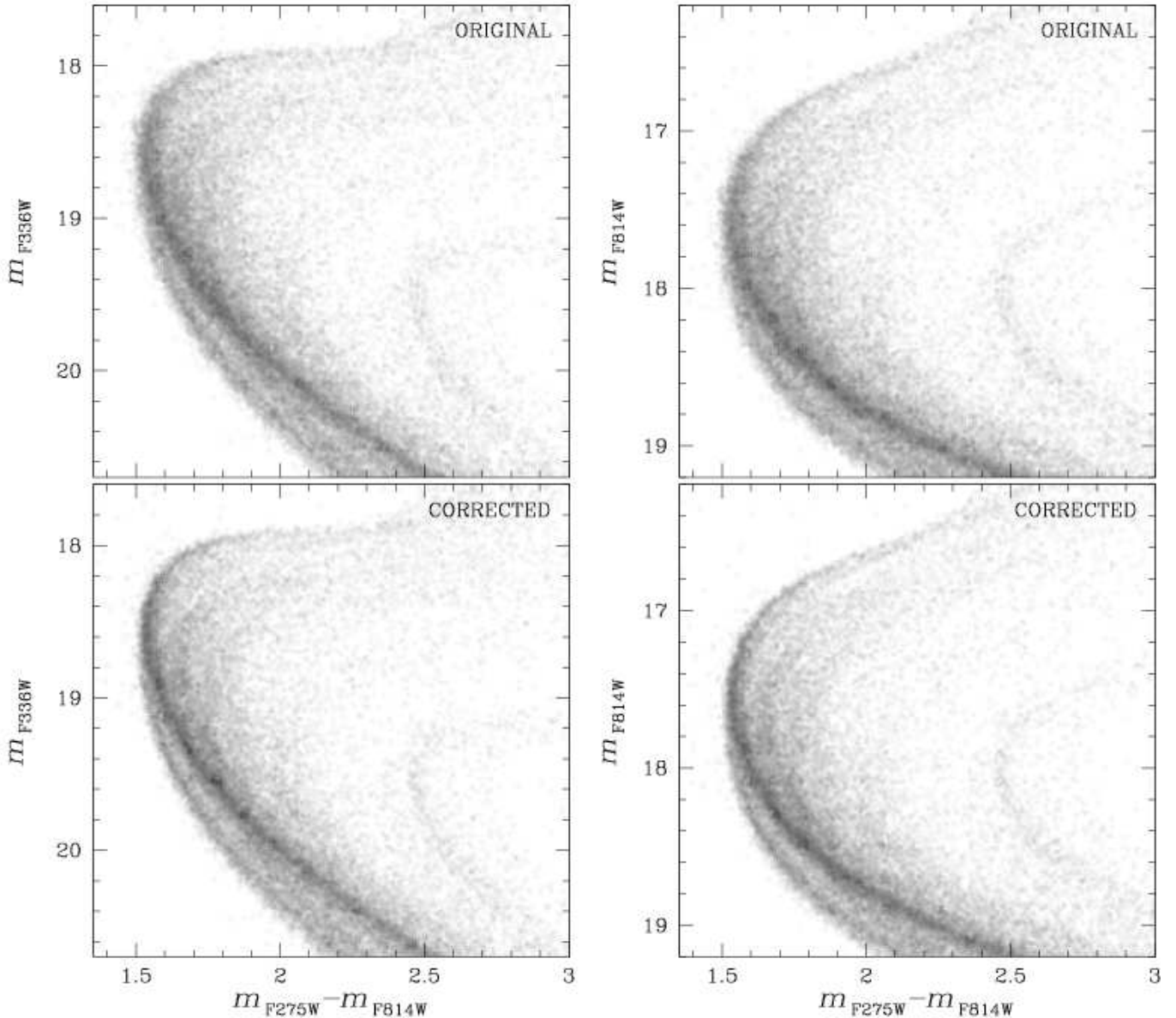}
\caption{\small{Top panels:\ the $m_{\rm F336W}$ vs.\ $m_{\rm
      F275W}-m_{\rm F814W}$ (left) and $m_{\rm F814W}$ vs.\ $m_{\rm
      F275W}-m_{\rm F814W}$ (right) Hess diagrams made using
    the original CMDs, not corrected for DR.  Bottom
    panels:\ Hess diagrams of the same CMDs once our
    DR-correction is applied.}}
\label{f:ba}
\end{figure*}

\begin{figure*}[!t]
\centering
\includegraphics[width=15.cm]{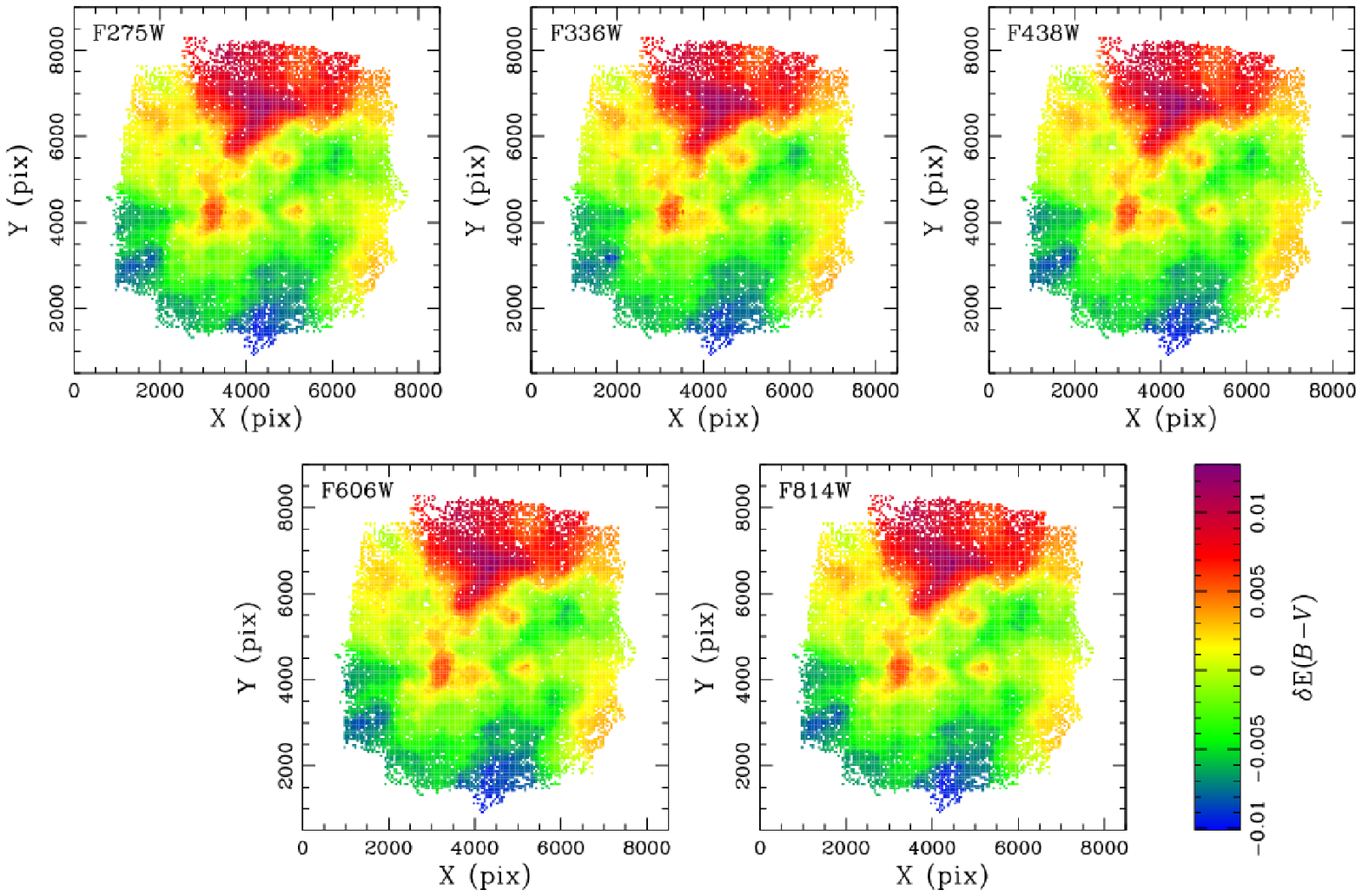}
\caption{\small{The five independent DR maps we obtained, color-coded
    using the same color-mapping scheme (shown in the bottom-right in
    the figure). The similarity of these maps is remarkable, both in
    terms of shape and intensity of the computed DR. The cluster
    center is approximately located at pixel position $(4300,4990)$.}}
\label{f:maps}
\end{figure*}

In Paper~I, we measured stellar photometry through three different
methods. Method one works best for bright, unsaturated stars, which
are able to produce a distinct peak within its local
neighbor-subtracted 5$\times$5-pixel raster in each individual
exposure.  Fainter stars often do not produce a significant peak in
each exposure, and their photometry is measured best using method two
and method three. Method-two starts from the stellar positions
measured during the finding stage and determine a best PSF-fit
estimate of the stellar flux from the inner 3$\times$3 pixels. Method
three is superior to the other two methods only at the very faint
regime;\ it takes the flux of the brightest four pixels of a stellar
profile simultaneously from all exposures and weights it according
to the expected values of the PSF in those pixels. (We refer the
interested reader to Paper~I for a detailed description of the
data-reduction processes.)

In what follows we will make use of the method-two photometry. At the
SGB level, method-one and method-two photometric measurements are
comparable in quality. Method-one photometry is more accurate than
that of method two along the red-giant branch (RGB), but the opposite
is true along the main sequence (MS). In each of the five CMDs that we
used to estimate the amount of DR in the core of the cluster, the
angle of incidence of the reddening vector with respect to the bright
MS and the base of the RGB is still close enough to be orthogonal that
we can also take full advantage of stars in these evolutionary phases to
correct for DR. Since MS stars are far more plentiful than RGB stars,
method-two provides the best photometry overall.

For all five filters, we required a star to have at least two good
measurements ($N_{\rm g}$$\geq$2) that are at least 2.5 sigmas above
the local sky-background noise. This choice is motivated by the fact
that we want to correct DR across the widest possible FoV, but at the
same time we want to keep under control the number of poorly-measured
stars near the edges of the FoV and minimize false detections (e.g.,
cosmic rays). We also excluded all those stars for which the $o$
parameter\footnote[3]{The $o$ parameter tells us the fraction of light
  present in the PSF fitting radius due to neighbors before neighbor
  subtraction.} is larger that 1. Then, separately for each filter, we
applied two cuts based on the photometric RMS and the \texttt{QFIT}
parameter\footnote[4]{The \texttt{QFIT} parameter tells us how well a
  stellar profile is fitted by the PSF.}  (Fig.~\ref{f:sel}). To do
this, we binned the stars every 0.1 magnitude and computed the
85$^{\rm th}$ percentile of the photometric RMS and \texttt{QFIT}
distributions in each bin. We set three hard constraints:\ stars are
always well-measured if their \texttt{QFIT} values are above 0.95, and
are always rejected if their \texttt{QFIT} values are below 0.4 or if
their photometric RMS is larger than 0.25
magnitudes.\footnote[5]{These fixed, arbitrary cuts will have no
  effects for the purpose of the present analysis, but we report them
  anyway for completeness. The same selection criteria will be applied
  for the analysis of the multiple stellar populations along the MS,
  that will be presented in the next paper of this series.}. We then
fitted a third-order polynomial to the computed percentile
values as a function of the magnitude (red lines in panels a1-2, b1-2,
c1-2, d1-2 and e1-2 of Fig.~\ref{f:sel}) on both the \texttt{QFIT}
vs.\ magnitude and photometric RMS vs.\ magnitude
planes.\footnote{Alternatively, more advanced techniques
    could have been used here, e.g., fitting a cubic spline to the
    85\% quantile of the two-dimensional point distributions, which
    falls into the category of nonparametric density estimations, or
    other very capable techniques of local regression like, e.g., the
    Nadaraya-Watson kernel regression (\citealt{nada64, watson64}),
    the LOESS model (\citealt{cle79}), spline regression (e.g.,
    \citealt{silv84}) or (for stationary Gaussian scatter) Gaussian
    Processes regression (`kriging', \citealt{krige51}).}  These
lines separate the bulk of well-measured stars from the few clear
outliers in each panel.  84$\,$155 stars appeared to be good according
to all the selection criteria.

Finally, we required selected stars to be cluster members. Of the
84$\,$155 high-photometric-quality stars, only 72$\,$657 have a
measured proper motion. The vast majority of 11$,$498 stars with no
proper-motion measurements are close to the faint magnitude limit (see
also Paper~I for a description of the proper-motion catalog).  In
panel (f) of Fig.~\ref{f:sel} we show the proper-motion diagram of the
72$\,$657 high-photometric-quality stars. On this panel we defined as
cluster members all those stars within the red circle (which radius is
arbitrarily defined as 4.2 times the average proper-motion
dispersion). The 48 stars outside the red circle, marked with red
crosses, are likely field stars.

The final sample contains 72$\,$609 proper-motion selected,
well-measured stars in all five filters, homogeneously displaced in a
roughly square region of about $4\farcm3$$\times$$4\farcm3$ around
the cluster's center.

\subsection{Reference Stars}
\label{ss:ref}

As mentioned above, we prefer to use as reference stars a large number
of objects belonging to a single stellar population, and in a
region of the CMD where the sequence of this population is as
perpendicular as possible to the direction of the reddening vector.

The most-populated sequence of \wcen\ around the SGB region is easily
recognizable as the brighter branch on the $m_{\rm F814W}$
vs.\ $m_{\rm F275W}-m_{\rm F814W}$ CMD (panel (a) of
Fig.~\ref{f:ref}).  Stars belonging to this sequence are the progeny
of red MS stars.  The red arrow (here and in the other panels of the
figure) marks the direction of the reddening vector, which is almost
perpendicular to the turn-off/SGB regions. Clearly, stars around the
turn-off, SGB and upper MS regions are the best reference-star
candidates.

On this CMD we initially selected our reference-star sample by hand
(panel (b), red outline), by choosing all objects likely belonging to
the most-populated sequence.  We limited reference stars to be within
the magnitude range 16$\leq$$m_{\rm F814W}$$\leq$18.8.  At fainter
magnitudes, the MS of this population bends progressively to become
almost parallel to the reddening vector, and therefore fainter MS stars
do not help us in correcting for DR effects.  On the other hand, the
inclusion of brighter stars would have made little to no difference,
given the scarcity of RGB stars with respect to the much more numerous
MS stars.

Now, it is well known that the different populations in \wcen\ overlap
each other around the turn-off region, with the red MS mostly evolving
into the bright SGB and bluer RGB branches, while the progeny of the
blue MS mostly populates fainter SGB and redder RGB branches (see,
e.g., \citealt{2010AJ....140..631B}). We can clearly notice the
blue-MS/red-MS flipping around the turn-off region also in panel (a)
of Fig.~\ref{f:ref}. As a result, our preliminary reference-star
selection necessarily includes objects actually belonging to
different stellar populations.

In \cite{2010AJ....140..631B} we showed that the blue and the red MSs
switch their relative position on a CMD based on the $m_{\rm
  F336W}-m_{\rm F438W}$ color. Panel (c) of Fig.~\ref{f:ref} shows
the $m_{\rm F814W}$ vs.\ $m_{\rm F336W}-m_{\rm F438W}$ CMD around the
turn-off region for the stars selected in panel (b). It is clear from
the figure that our preliminary selection encompasses more than one
population of stars, especially below the turn-off level ($m_{\rm
  F814W}$$\sim$17.5).  We improved our reference-star sample (panel
d) by keeping only those stars (in azure) within the red outline
drawn by hand.

A small fraction of contaminants might still be present in our
reference-star sample (e.g., stars belonging to different
subpopulations).  We explored the possibility of further refining our
list of reference stars by plotting them in all the CMDs and two-color
diagrams made available by our five adopted filters.  The reference
stars selected in panel (d) of Fig.~\ref{f:ref} define a single
sequence in all cases, without any obvious hint of a split or clear
contamination from other populations. Therefore, we considered our
reference-star list defined in panel (d) as the final sample, which
contains 11$\,$655 stars.

\subsection{Choosing the Appropriate Correction Plane}
\label{ss:plane}

Because the reddening vector's direction is diagonal with respect to
the axes in a CMD, both the magnitude and the color of stars have to
be corrected for the effects of DR.  If we rotate the CMD in such a
way that the reddening vector is parallel to one of the axes, then we
have to correct only along one direction, thus simplifying the
process. As an example, in panel (a) of Fig.~\ref{f:how} we show the
$m_{\rm F438W}$ vs.\ $m_{\rm F275W}-m_{\rm F814W}$ CMD, in which
reference stars are highlighted in azure and the reddening vector is
marked in red. We chose an arbitrary rotation point (red encircled dot),
and we rotated the CMD counterclockwise by the angle
$$\alpha=\arctan\left(\frac{{\rm A}_{\rm F438W}}{{\rm A}_{\rm
    F275W}-{\rm A}_{\rm F814W}}\right).$$ The resulting rotated
pseudo-CMD is in panel (b), zoomed-in around the SGB. We call the X
and Y coordinates of this plane as ``col'' and ``mag'',
respectively. For completeness, panel (b) also shows the rotated
direction of the reddening vector (red arrow), which is now parallel
to the ``col'' axis, as well as the location of the rotation point.

To further simplify the DR-correction process, we rectified the
pseudo-CMD as follows. We computed a fiducial line for the
reference-star sequence by least-squares fitting a 5th order
polynomial to the ``col'' and ``mag'' values of its stars (red curve
in panel b). Then, we subtracted from the pseudo-color of each star
that of the fiducial line at the same ``mag'' level. With a rectified
reference-star sequence, we can better monitor the following steps of
the DR-correction process. The resulting rectified pseudo-CMD is shown
in panel (c). Let us call $\Delta$``col'' the abscissa of this panel.
In order to have an even distribution of reference stars around the
fiducial line at any given ``mag'' level --and therefore avoid any
correction bias at the faint and/or the bright end during the next
correction steps-- we excluded those reference stars (in gray) outside
the two red horizontal lines in panel (b).

On a side note, it is worth noting that most of the other SGB branches
are tilted with respect to the direction of the rectified
reference-star fiducial sequence (panel c). Most of the stars in
tilted SGBs that overlap the reference-star sequence are the progeny
of the blue MS. The ``tilt'' is due to a difference in chemical
abundances between blue and red MS stars. It is also worth noting that
some SGB sequences are more tilted than others.  We will return and
focus on the multiple stellar populations along the SGB of \wcen\ in a
future paper of this series.

\subsection{Local Corrections}
\label{ss:loc}

For each star in the catalog (hereafter, the \textit{target} star), we
identify the spatially-closest 75 reference stars, which are used to
compute their 2.5-$\sigma$-clipped $\Delta$``col'' median
value.\footnote[6]{The \textit{target} itself is not included if it
  happens to also be a reference star.}  If the \textit{target} star
also happens to be in the reference list, then the star itself is not
used to compute the $\Delta$``col'' median value.  Panel (d) of
Fig.~\ref{f:how} shows the outline of the FoV of our data set, in
which we marked the position of two \textit{target} stars:\ A and B. The
closest 75 reference stars to A and B are found within the two red
circles of radius 254.6 and 321.2 pixels, respectively (or about
10$\farcs$2 and 12$\farcs$8).

Panels (e) and (f) of Fig.~\ref{f:how} show the $\Delta$``col'' values
of the closest 75 reference stars to A and B, respectively. The
2.5$\sigma$-clipped median $\Delta$``col'' value in both panels is
marked by a red vertical line, while red crosses are rejected stars.
The DR correction (for the rotated pseudo-CMD) to be applied to \textit{target}
stars A and B is the opposite of these median $\Delta$``col'' (red
arrows in panels e and f).  The error associated to the DR correction
of stars A and B is given by the error of the mean of the associated
reference-star median $\Delta$``col'' values. For both \textit{target} stars in
the example, we have a correction error of about 0.0042
$\Delta$``col'' magnitudes (or about 10\% and 6\% the applied
correction for A and B, respectively), which translates into a CMD color
and magnitude error of about 0.003 magnitudes.

Once we have corrected all stars in the catalog on the rotated
pseudo-CMD, we can rotate the pseudo-CMD back by the angle $-\alpha$,
to obtain the properly DR-corrected CMD.

\subsection{Iteration}
\label{ss:iter}

The DR-corrected CMDs that we have created allow us to fine-tune our
original reference-star selections, because on them it is easier to
discriminate between stars belonging to the most-populated sequence
from the other populations.

We started from the same color-magnitude planes shown in
Fig.~\ref{f:ref}, this time using DR-corrected photometry, and we
improved the selection of reference stars by slightly narrowing the
color width of the boxes shown in Fig.~\ref{f:ref}. Moreover, thanks
to the DR-corrected photometry, we realized that the selected
reference-star sequence splits into 2 branches for magnitudes $m_{\rm
  F814W}<16.5$ (see, e.g., panel d of Fig.~\ref{f:ref} for a reference),
therefore we limited reference stars to be fainter than this
brightness limit.  The improved reference-star list contains 9051
objects.

At this point, we started an iteration process that consists of two
steps:\ 1) we repeated the DR-correction procedures described in the
previous subsection; and 2) we improved the list of reference stars
using the resulting DR-corrected CMDs. During the iteration
procedures, we kept fixed the selection boxes used to define the
reference-star sample.  Note that the DR-correction is always computed
using the raw, uncorrected photometry. What changes from one iteration to
the other is the number of reference stars.

After the third iteration, the sequences on the DR-corrected CMDs were
negligibly narrower than those after the second iteration, and the
number of improved reference stars for an additional, fourth iteration
changed only by less than 0.1\%. Thus, we did not iterate the
DR-correction further, and we considered the results of the
third iteration as our final correction.

In Fig.~\ref{f:ba} we show the Hess diagrams of two of the
five CMDs used in this analysis around the SGB region before (top) and
after (bottom) the DR-correction is applied, as a demonstration of the
effectiveness of the method. The Hess diagrams based on the
$m_{\rm F336W}$ vs.\ $m_{\rm F275W}-m_{\rm F814W}$ CMDs are on the
left, the Hess diagrams based on the $m_{\rm F814W}$
vs.\ $m_{\rm F275W}-m_{\rm F814W}$ CMDs are on the right. The
multiple-population sequences in the bottom panels are narrower after
the correction, and finer details of SGB and MS substructures are now
visible.

\begin{figure*}[!t]
\centering
\includegraphics[width=\textwidth]{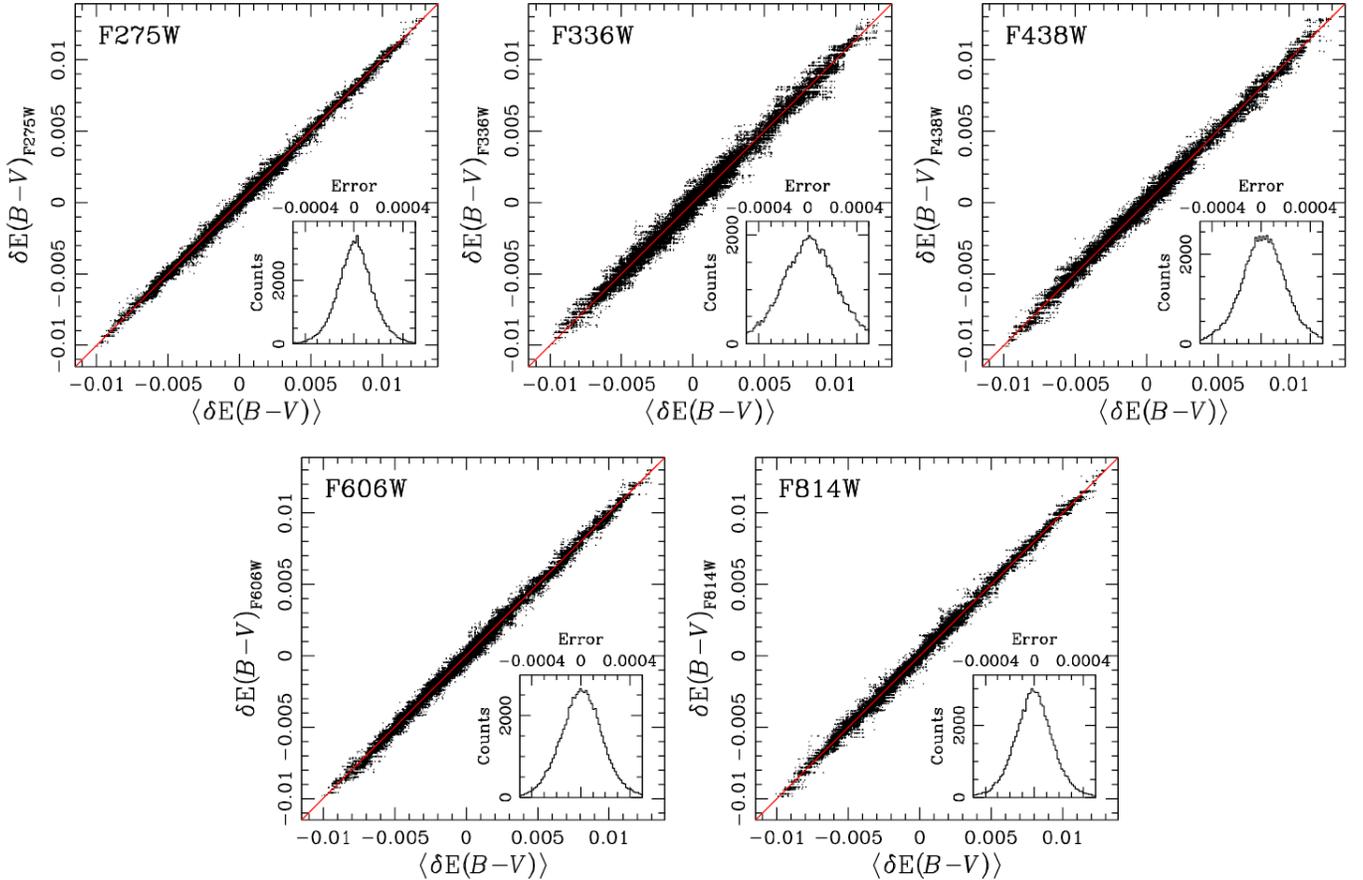}
\caption{\small{Star-to-star comparison of the filter-based
    DR-correction values versus the median DR-correction.
    The red line in each panel is the bisector of the plane (1:1
    correlation) and not a fit to the points. The inset in each panel
    shows the distribution of the points in the direction orthogonal to
    the red line, which provides an estimate of the internal errors.}}
\label{f:ana}
\end{figure*}

\section{Reddening-Map Analysis}
\label{s:maps}

One of the by-products of our DR-correction procedures is a DR map of
the field.  As briefly mentioned at the beginning of the previous
Section, in principle our corrections account for both DR and
photometric zero-point variations across the FoV.  If the contribution
of the latter is significant, then we should expect sizable
differences between DR maps obtained from different CMDs.  If there
are marginal variations. that would imply that the high-precision PSF
models we used are also highly accurate.  The comparison of DR maps
based on different CMDs thus represents a powerful diagnostic tool for
the characterization of the quality of our PSF models.

The DR maps based on the 5 CMDs are constructed as follows. First, we
considered only the component of the DR correction along the Y axis,
which is the only axis that changes from one CMD to another (being the
$m_{\rm F275W}-m_{\rm F814W}$ color fixed in all five CMDs). Let us
consider the $m_{\rm F275W}$ vs.\ $m_{\rm F275W}-m_{\rm F814W}$ CMD as
a guideline. For a given star, its DR-corrected $m_{\rm F275W}^{\rm
  CORR}$ magnitude is obtained as:\
$$
m_{\rm F275W}^{\rm CORR}=m_{\rm F275W}^{\rm RAW}-\frac{{\rm A}_{\rm
    F275W}}{{\rm E}(B-V)}\delta {\rm E}(B-V)_{\rm F275W},
$$ 
where $\delta {\rm E}(B-V)_{\rm F275W}$ is the applied DR value in
magnitudes, and $m_{\rm F275W}^{\rm RAW}$ is the stellar raw,
uncorrected magnitude.  Using the appropriate extinction-coefficient
value for the F275W filter listed in Table~1, it follows:\
$$
\delta {\rm E}(B-V)_{\rm F275W}=\frac{m_{\rm F275W}^{\rm RAW}-m_{\rm
    F275W}^{\rm CORR}}{6.10379}.
$$ 
For each star we can compute five independent DR values, one for each
of the five filters.

The values $\delta {\rm E}(B-V)_{\rm filter}$ as a function of the
(X,Y) stellar position can be used to construct two-dimensional maps
of the DR across the FoV.  The five panels in Fig.~\ref{f:maps} show
 these maps, one for each of the five filters (listed
in the top-left corner of each panel).  The panels adopt the same
color mapping, the scale of which is shown in the bottom-right part of
the figure. These five maps look remarkably similar to each other,
both in shape and intensity. This means that our carefully-modeled
PSFs (see Paper~I, Section~3.1), together with the availability of a
large number of dithered exposures with different roll angles, were
able to minimize the effects of photometric zero-point residuals.

For each star we compared the five $\delta {\rm E}(B-V)_{\rm filter}$
values we obtained with respect to their median value $\langle \delta
{\rm E}(B-V)\rangle$, which represents our best estimate of the true
DR for that star (Fig.~\ref{f:ana}). The red line in each panel is the
bisector of the plane and not a fit to the points. Each panel reveals
the presence of small systematic deviations from the 1:1 correlation,
especially at the extremes of distribution, and in particular for the
$\delta {\rm E}(B-V)_{\rm F336W}$ and $\delta {\rm E}(B-V)_{\rm
  F438W}$ values.  Nonetheless, the fact that in all cases the points
are nicely aligned along the red line represents a clear validation of
our technique. We least-squares-fitted a straight line to the
unweighted points in each panel, and the slopes we obtained are indeed
very close to the value of 1 (see Table~1, third column).

\begin{table}[!th]
\label{tab}
\centering
\small{
\begin{tabular}{cccc}
\multicolumn{4}{c}{\textsc{Table 1}}\\
\multicolumn{4}{c}{\textsc{Filter-based DR Comparison}}\\
\hline\hline
\textbf{Filter}&\textbf{A$_{\rm filter}$/E($B-V$)}&\textbf{Slope}&
\textbf{$\sigma$}\\
\hline
F275W & 6.10379 & 1.00102$\pm$0.00018 & 0.00014\\
F336W & 5.04267 & 0.99853$\pm$0.00036 & 0.00022\\
F438W & 4.04616 & 0.99502$\pm$0.00027 & 0.00018\\
F606W & 2.62074 & 1.00091$\pm$0.00023 & 0.00017\\
F814W & 1.64912 & 1.00089$\pm$0.00010 & 0.00015\\
\hline\hline
\end{tabular}}
\end{table}

In each panel, the dispersion of the points orthogonal to the red line
gives us an independent estimate of the internal errors. A histogram
of the distribution of these orthogonal dispersions is shown in the
inset of each panel. A Gaussian fit to these histograms provides a
first-guess estimate of the internal errors, which are found to be
smaller than 2$\times$10$^{-4}$ mag in all five cases. The $\sigma$ of
the fitted Gaussians are reported in the fourth column of Table~1.  A
somewhat larger $\sigma$ for the F336W filter might suggest that our
PSF models for this filter were not as fine-tuned as for the other
filters.

\begin{figure*}[!t]
\centering
\includegraphics[width=17.3cm]{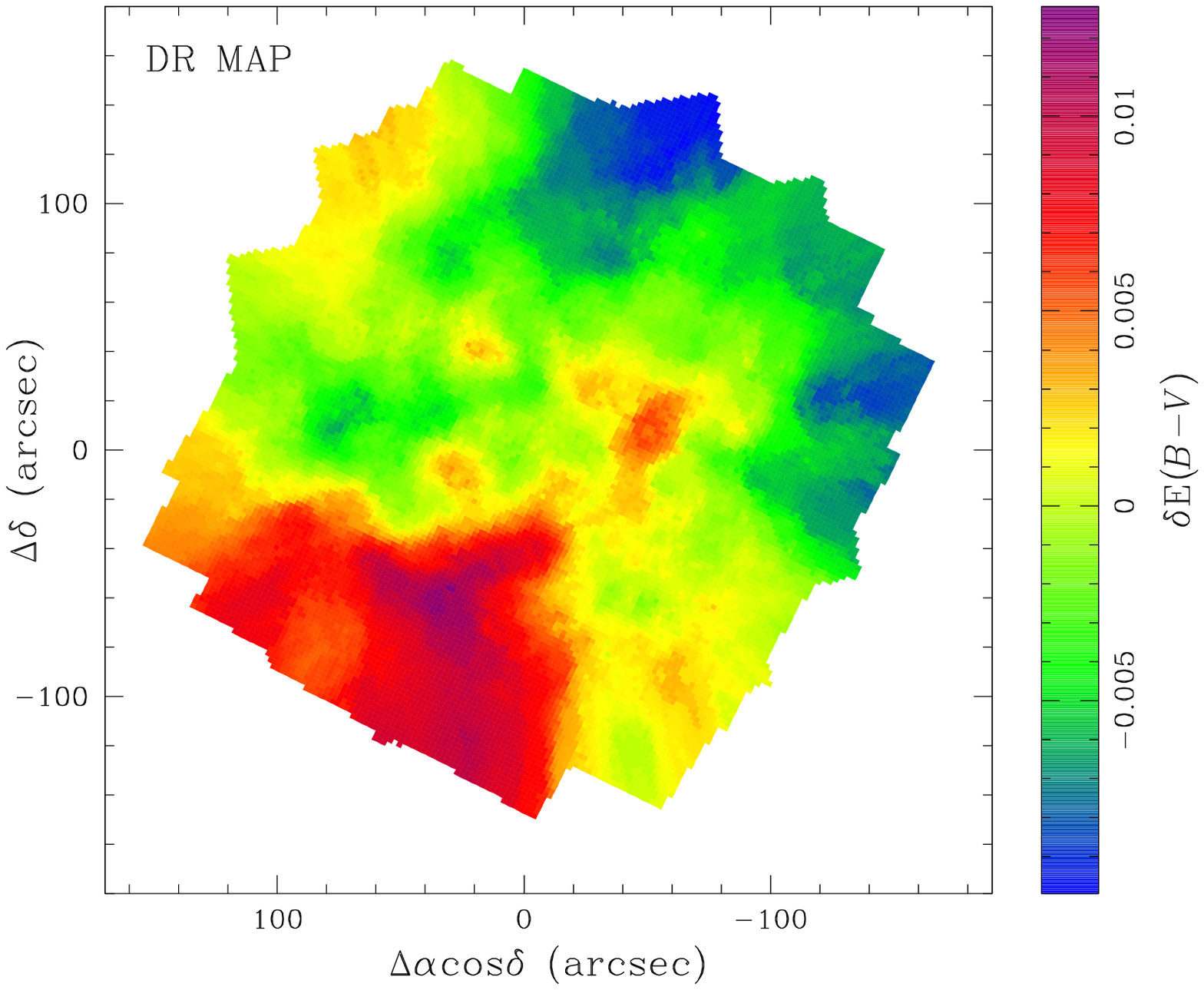}\\
\includegraphics[width=\columnwidth]{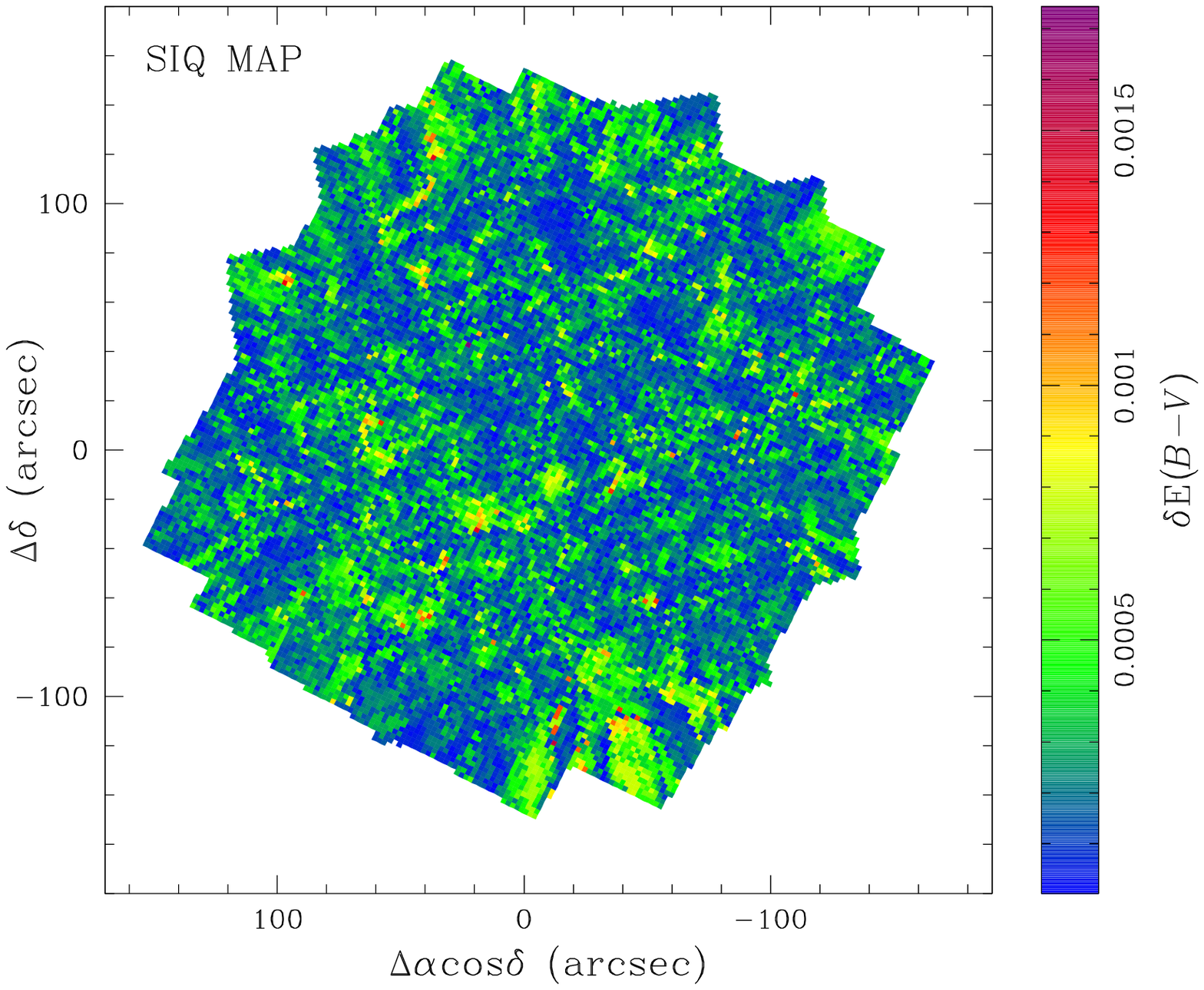}
\includegraphics[width=\columnwidth]{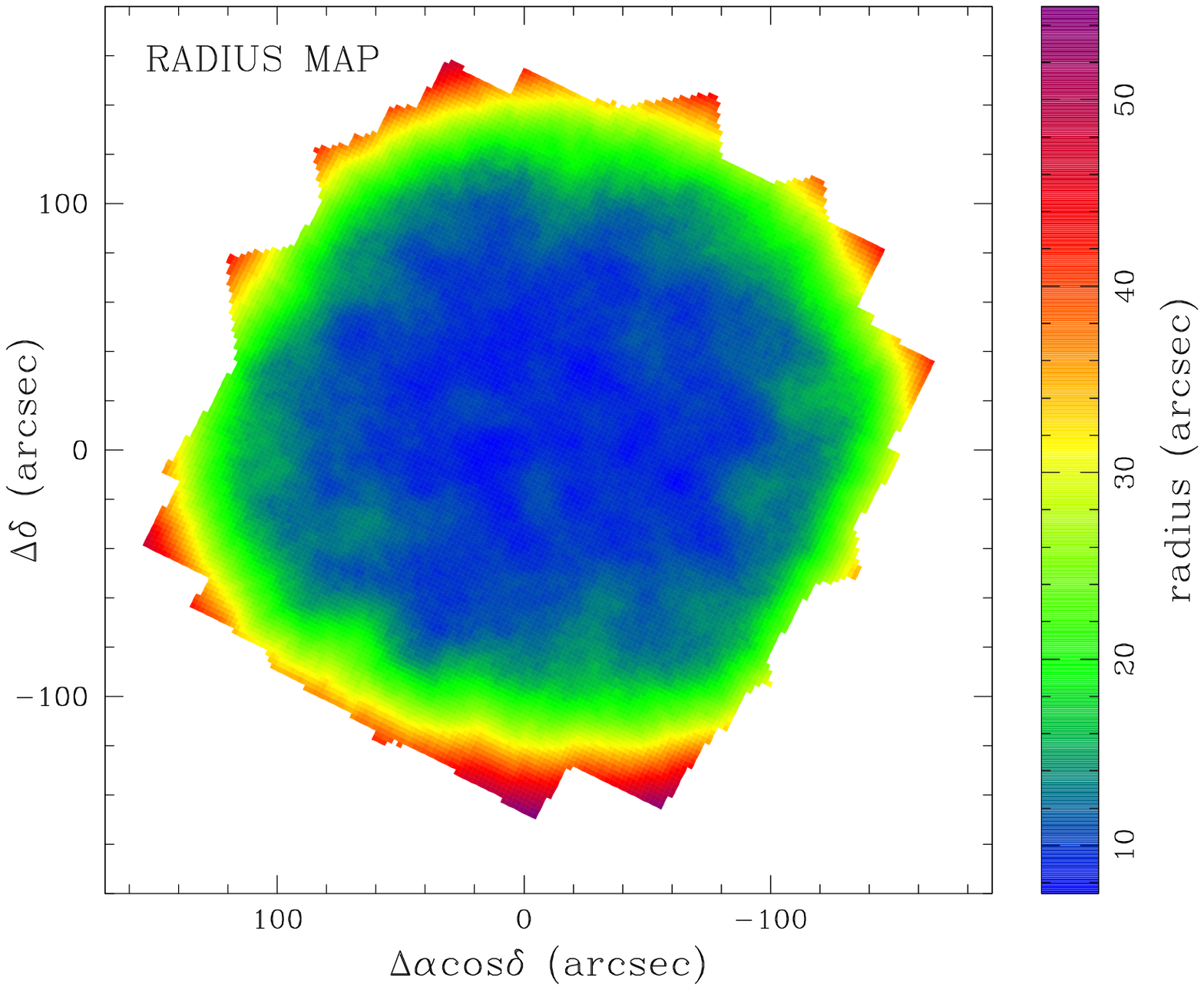}
\caption{\small{Top panel:\ the final DR map of the core of \wcen,
    shown in the rectified Cartesian plane with respect to the
    cluster's center. Bottom left:\ The local semi-interquartile (SIQ)
    values of the five filter-dependent DR maps across the FoV. Bottom
    right:\ a map of how far from the center of each 50$\times$50
    pixels$^2$ 75 reference stars are found (1 pixel = 40 mas). The
    color-coding of each panel is shown to its right side.}}
\label{f:fits}
\end{figure*}

Now, a more careful look at the values of the slopes we obtain reveals
that some of these values are significantly different from 1
(especially for the F438W filter), even if we take into account both
the error of the slope and the $\sigma$ of the Gaussian fit.  At
first, we thought that we could take advantage of these results and
attempt to recover both the true DR map and the true extinction curve
from our data set. We tried to iterate the DR-correction procedures by
adjusting each time the extinction coefficients in such a way as to
have the five slopes in Fig.~\ref{f:ana} identically equal to
1:\ A$_{\rm filter}^{\rm new}$=A$_{\rm filter}^{\rm orig}/{\rm
  slope}$, but the improved values of the slopes that we obtained
after each iteration kept oscillating around their initial values and
never converged to unity as we would have hoped.

Then, we realized that we are actually facing a double-degenerate
problem. For each of the five CMDs, the only things we know
observationally are the location of the reference-star fiducial line
and the location of the \textit{target} stars. We do not know the true value of
the extinction coefficients, but only their estimate. In addition, we
do not know the true DR affecting each \textit{target} star, but only an
estimate. Finally, we also do not know the true position of the
reference stars on the $\Delta$``col'' plane as they would appear at
the average reddening of the cluster, but only that these stars must
be on the fiducial line. For a given reference star, there are a total
of 7 unknown quantities, namely:\ its true DR, its true position on
the $\Delta$``col'' plane, and the true value of the extinction
coefficients. 

Regardless of the number of filters $M$ we use to solve for the DR,
there will always be $M+2$ unknowns. Hence, there is a 2-parameter
family of solutions.  One degeneracy arises from the fact that only
combinations of A$_{\rm filter}$$\times$$\delta$E($B$$-$$V$)$_{\rm
  filter}$ are observable, therefore there is a multiplicative
degeneracy between the true DR values $\delta$E($B$$-$$V$)$_{\rm
  filter}$ and the extinction coefficients A$_{\rm filter}$. A second
degeneracy arises from the fact that we do not know where a reference
star really lies with respect to the reference-star fiducial line.

Given these degeneracies, we cannot determine the DR and the
extinction curve simultaneously. The approach we take in this paper is
to assume that we know the value of the extinction coefficients, and
we then determine the DR. This approach yields a unique solution,
which is in fact what we encounter. The slope values in Table~1
indicate that different CMDs imply a (slightly) different differential
reddening.  This means that the adopted \cite{1999PASP..111...63F}
extinction curve is not consistent with our data at the $\sim$0.1\%
level.  However, the fact that the slope values in Table~1 are so
close to unity indicates that the uncertainty in our DR maps due to
uncertainties in the extinction coefficients are very small.  In
principle, the results can be improved by changing the assumed
extinction curve. However, there is no unique way of doing this, since
there is a two-parameter family of solutions. Hence, an iterative
approach such as the one we tried does not converge.

\subsection{The Final Differential-Reddening Map}
\label{ss:final}

The top panel of Fig.~\ref{f:fits} shows our final, $\langle \delta
{\rm E}(B-V)\rangle$ map, in the rectified Cartesian (R.A.,Dec.)
plane, in units of arcseconds with respect to the cluster's center.
This map represents our best estimate of the real DR in the field. To
create the map, we sampled the available FoV every 50$\times$50
pixel$^2$ (i.e., 2$\times$2 arcsec$^2$), and we applied to each of
these 2500 pixels the median value of the five filter-dependent DR
corrections computed at the center of the cell.  A
50$\times$50-pixel$^2$ sampling provides for a smooth representation
of the DR map, and largely supersamples the typical resolution of our
DR-correction (given that, on average the closest 75 reference stars
to a given \textit{target} star are found within 330 pixels, or
13$^{\prime\prime}$).

The map has an average value of zero across the entire field (the size
of which is approximately $4\farcm3$$\times$$4\farcm3$), since we
computed the DR with respect to the average reddening of the field.

Overall, the DR values on the map have a standard deviation of 0.005
${\rm E}(B-V)$ magnitudes. Minimum and maximum ${\rm E}(B-V)$ values
are $-$0.010 and +0.013 magnitudes, respectively, which (assuming an
average ${\rm E}(B-V)$=0.12 magnitudes) translate into a typical DR
variation of 4\%, and up to about $\pm$10\% (at most) the average
reddening. Another way to read these results is that the DR in the
core of \wcen\ can vary by up to 20\% (minimum to maximum).

The bottom-left panel of the figure shows a map of the
semi-interquartile (SIQ) of the five filter-based $\delta {\rm
  E}(B-V)$ values at each of the 50$\times$50 pixel$^2$ locations,
shown in the same rectified Cartesian plane as the top-panel map.  The
SIQ provides an estimate of the local scatter between the five DR
estimates. Overall, the SIQ-value distribution is flat across the FoV,
with typical variations of the order of 0.0003 ${\rm E}(B-V)$ (less
than 10\% the DR standard deviation).  There are a few localized
regions, mostly close to the edges of the FoV where we have only a few
(typically two) exposures per filter, or in close proximity of
oversaturated stars in some filters, in which the SIQ is as high as
0.0016 ${\rm E}(B-V)$.

Finally, the bottom-right panel shows the rectified Cartesian map of
the typical values of the radius within which 75 reference stars are
found for each of the 50$\times$50 pixel$^2$ locations. The
color-mapping scale is in pixel units. Given the WFC3/UVIS pixel scale
of 40 mas pixel$^{-1}$, the minimum and maximum values are
10$^{\prime\prime}$ and 43$^{\prime\prime}$, respectively.  Clearly,
the DR map is less constrained near the corners of the FoV, where 75
reference stars are found only along one direction (toward the center
of the FoV).

\begin{figure}[!t]
\centering
\includegraphics[angle=90,width=\columnwidth]{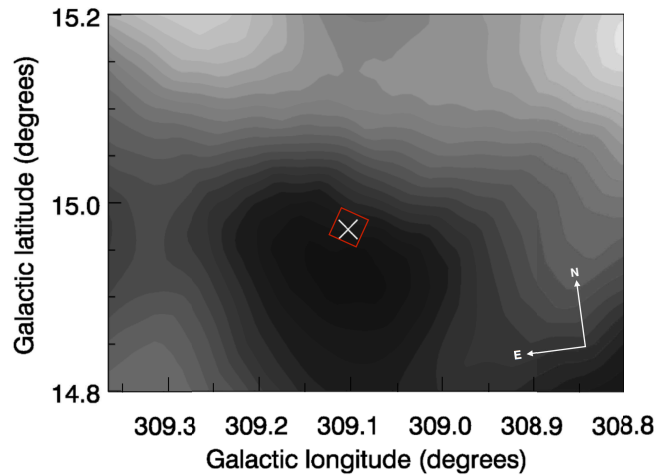}
\caption{\small{The DIRBE/IRAS reddening map in the direction of
    \wcen\ (\citealt{1998ApJ...500..525S}), adapted from \cite[their
      Fig.~10b]{2009MNRAS.399..195V}. The cluster's center is marked
    by a white cross. The outline of our FoV is in red. The gray-scale
    intensity of the map ranges from E($B-V$)=0.144 mag (dark)
    to 0.124 mag (light).}}
\label{f:vl}
\end{figure}

\section{Discussion and Conclusions}
\label{s:disc}

In this paper, we took advantage of the exquisite astro-photometric
catalog we constructed in Paper~I to derive a high-precision DR map of
the core of the globular cluster \wcen. The DR is estimated using five
CMDs based on the following UVIS filters:\ F275W, F336W, F438W, F606W
and F814W.  There are at least 34 single exposures taken
through each of these filters, with different roll angles and dither
patterns, so that the impact of PSF-related, zero-point spatial
variations across the field is minimized to the extent possible.  In
addition, images taken through these filters cover the widest
available FoV around the cluster's core. 

The five independent solutions for the DR we found agree very well
with each other, with typical semi-interquartile values of 0.0003
${\rm E}(B-V)$.  The median of the five DR maps constitutes our best
estimate of the true DR around the core of the cluster. Within our FoV
(about $4\farcm3$$\times$$4\farcm3$) the DR has standard deviation,
minimum and maximum values of 0.005 mag, $-$0.010 mag and 0.013 mag in
terms of ${\rm E}(B-V)$, respectively. Assuming an average reddening
of \ebv=0.12 (\citealt{1996AJ....112.1487H}), these values translate
into a typical DR variation of about only 4\%, and up to
$\sim$$\pm$10\% at the extremes.

Our findings are consistent with the qualitative analysis of
\cite{2007ApJ...663..296V}, but are in sharp contrast with the results
of \cite{2005ApJ...634L..69C}, who found clumpy reddening variations
of almost a factor of two (minimum to maximum) within the cluster's
core.  In addition, \cite{2005ApJ...634L..69C} found that the greatest
density of more highly-reddened objects is shifted along the right
ascension axis when compared with less reddened ones. Our DR map, on
the other hand (top panel of Fig.~\ref{f:fits}), shows that the
regions with the highest and lowest DR are the South-East and the
North-West quadrants, respectively. Both direction and orientation of
the DR gradient we find are qualitatively consistent with the
DIRBE/IRAS reddening map of the region compiled by
\cite{1998ApJ...500..525S}. In Fig.~\ref{f:vl} we show a portion of
this DIRBE/IRAS map, adapted from \citet[their
  Fig.~10]{2009MNRAS.399..195V}. The footprint of our FoV is
highlighted in red, while the cluster's center is marked by a white
cross. The \cite{1998ApJ...500..525S} DR map, despite having a much
lower spatial resolution than ours, clearly suggests a DR gradient
(high to low) from the South-East to the North-West with respect to
the cluster's center within our FoV.

A possible explanation of the differences between the
\cite{2005ApJ...634L..69C} DR analysis and both the DIRBE/IRAS DR map
and the DR map presented here is that \cite{2005ApJ...634L..69C}
treated the horizontal branch of \wcen\ as if it were the result of
the evolution of a ``single stellar population'', i.e., stars born at
the same time and with the same chemical composition. It has been
known for a long time that \wcen\ hosts stars with a wide range in
metallicity (e.g., \citealt{1973MNRAS.162..207C}), and more recently
that it also hosts distinct stellar populations in all evolutionary
sequences (e.g, \citealt{anderson97, 1999Natur.402...55L,
  2004ApJ...605L.125B}). More importantly, stars in \wcen\ exhibit a
large spread in helium content (e.g., \citealt{2004ApJ...612L..25N,
  2005ApJ...621..777P, 2012AJ....144....5K}):\ a spread that plays a
major role in shaping the blue-horizontal-branch morphology (see,
e.g., \citealt{2004ApJ...611..871D, 2008MNRAS.390..693D}).

We release to the astronomical community our high-resolution,
high-precision, and high-accuracy DR map in the form of a
multi-extension \texttt{FITS} file containing the three maps shown in
Fig.~\ref{f:fits}. The DR map itself is in the first extension.  The
second extension contains the SIQ map, while the values of the
local-averaged radius (in pixels) within which 75 reference stars are
found are in the third extension.

\acknowledgments \noindent \textbf{Acknowledgments.} AB acknowledges
support from STScI grants AR-12656 and AR-12845. GP acknowledges
partial support by PRIN-INAF 2014 e by the "Progetto di Ateneo 2014
CPDA141214 by Universit\`a di Padova.

\small{
}

\end{document}